\documentclass[aps,prd,twocolumn,showpacs,eqsecnum,amsmath,nofootinbib]{revtex4}
\usepackage[T1]{fontenc}
\usepackage[utf8]{inputenc}
\usepackage{lmodern}
\usepackage[english]{babel}
\usepackage{xcolor}
\usepackage{graphicx,amssymb,amsthm,amsmath,mathrsfs}
\usepackage{hyperref}

\def \d {{\rm d}}
\def \e {e}

\def \boldl {\mbox{\boldmath$l$}}

\begin{document}

\title{On the compatibility of nonlinear electrodynamics models with Robinson–Trautman geometry}

\author{T. Tahamtan}
\email{tahamtan@utf.mff.cuni.cz}

\affiliation{Institute of Theoretical Physics, Faculty of Mathematics and 
Physics, Charles University, Prague, V~Hole\v{s}ovi\v{c}k\'ach 2, 180~00 
Prague 8, Czech Republic}

\begin{abstract}
 Robinson--Trautman solutions with nonlinear electrodynamics are investigated for both  $\mathbb{L}(F)$ and $\mathbb{L}(F,G)$ Lagrangians and presence of electric and magnetic charges as well as electromagnetic radiation is assumed. Particular interest is devoted to models representing regular black holes for spherically symmetric situations. The results show clear uniqueness of Maxwell electrodynamics with respect to compatibility with the Robinson--Trautman class. Additionally, regular black hole models are clearly not suited to this class while the famous Born--Infeld model illustrates an important distinction between $\mathbb{L}(F)$ and $\mathbb{L}(F,G)$ as regards the electric field we obtain when the magnetic field is nontrivial.
\end{abstract}

\date{\today}

\maketitle

\section{Introduction}

Nonlinear Electrodynamics (NE) is a generalization of the linear Maxwell theory to a nonlinear theory.  Such theory was initially developed to solve the problem of the divergent field of a point charge (see, e.g. \cite{Dirac}) also yielding a reasonable self-energy of a charged particle. The best-known and frequently used model of NE was proposed by Born and Infeld in 1934 \cite{BornInfeld}. An excellent overview of nonlinear electrodynamics and its main features was given in a book by Pleba\'{n}ski \cite{Plebanski}. Einstein gravity coupled with nonlinear electrodynamics has attracted attention of many researchers in the literature and this carried over to modified theories as well (for example \cite{Tahamtan-fR}). Considerable interest is specifically devoted to gravitating NE models which resolve (remove) the spacetime singularity in the center of a black hole. Such solutions are called regular black holes and their association with NE started with the model proposed in \cite{Ayon-Beato1}.

Generally, the Lagrangian $\mathbb{L}$ of nonlinear electrodynamics is supposed to be a scalar function of the invariants $F=F_{\mu \nu}F^{\mu \nu}$ and $G=F_{\mu \nu}{}^{*}F^{\mu \nu}=\frac{1}{2}\epsilon^{\mu \nu \alpha \beta}F_{\mu \nu}F_{\alpha \beta}$ (in fact one should consider only $G^{2}$ to eliminate the pseudoscalar nature of $G$). If one applies the NE theories in static spherically symmetric spacetimes then the form of the Lagrangian reduces to $\mathbb{L}(F)$, which is also the one used most frequently in the literature. The $\mathbb{L}(F,G)$ form of the Lagrangian is used for studying light propagation in the geometric--optics approximation \cite{Novello2000}, particularly for comparisons with the linear Maxwell theory. For example in \cite{obukhov2002}, by analyzing Fresnel equations of wave propagation, the authors showed that there is no birefringence in the Born--Infeld model, but the velocity of light (as waves of NE) is different from $c$ and always less than or equal to $c$ (in Maxwell limit). Paper \cite{perlik2016} derived the conditions for causal propagation in $\mathbb{L}(F,G)$ theories and demonstrated that with the $\mathbb{L}(F)$ Lagrangian any theory other than the standard Maxwell vacuum one necessarily violates the causality conditions for some allowed background fields.

Apart from studies of the general physical properties of NE ($\mathbb{L}(F,G)$) there are not many nontrivial ($F\neq 0$, $G\neq 0$) exact solutions of Einstein gravity coupled with these models of nonlinear electrodynamics. One exception is a solution describing the Born--Infeld model in flat spacetime, which was obtained using the Newman--Penrose formalism already in \cite{bicak1975}.

This lack of solutions for the $\mathbb{L}(F,G)$--type model is the initial motivation for our research. Since this means stepping out of the spherically symmetric situation we decided to consider the Robinson--Trautman class, which essentially contains deformations of spherically symmetric situations. Additionally, this class contains generically dynamical solutions, which settle down to a symmetric situation by radiation of gravitational waves. This dynamical nature of the selected geometry has the additional benefit of making it potentially suitable for nonlinear stability studies of the NE solutions. Knowing whether a given static or stationary solution is stable or not yields another perspective of the spacetime. Unstable solutions usually are of less physical significance than stable ones since they are likely to decay to a stable configuration. 

However, one of the most frequent methods used to explore the stability of solutions is linear stability analysis. In \cite{Sarbach2003}, Moreno and Sarbach presented a study of dynamical stability of black--hole solutions in self-gravitating nonlinear electrodynamics with respect to arbitrary linear fluctuations of the metric and electromagnetic field. They established stability for several specific models of NE--- particularly those corresponding to regular black holes such as the Bardeen black hole--- based on some conditions on the electromagnetic Lagrangian. Based on the pulsation master equations obtained in \cite{Sarbach2003}, paper \cite{Sarbach-2016} obtained the fundamental Quasi-Normal (QN) modes associated with the gravitational and electrodynamic perturbations of black holes in NE theories. Parallel to this line of investigation there are several other studies of QN modes for NE \cite{QNmodes}  but their results apply only to test fields propagating in the fixed geometry of such black holes, unlike in \cite{Sarbach-2016}.

An alternative study of linear stability \cite{enegry-stability-2005} based on the sign of the effective energy shows that, except for the Born--Infeld model, most other models (and especially those for regular black holes) are unstable.

This disagreement concerning the linear stability established using different methods or imposing various conditions is exacerbated by an additional issue for regular black holes. Stability analyses are predominantly involving the region above the horizon. This is crucial for regular black holes since the ``removed'' singularity is below the horizon and the exterior solutions tend to be quite similar to other charged black hole solutions. It is thus important to specifically understand the stability of regular black holes below their horizon.

These issues might be resolved by finding exact solutions of Einstein gravity coupled to NE that would correspond to a dynamical spacetime with no symmetries and analysing their non--linear stability. Base on our past research \cite{Tahamtan-PRD-scalar, Tahamtan-2016-NE, wormehole}, the Robinson--Trautman (RT) spacetimes provide a potentially suitable candidate for such study.

Vacuum Robinson--Trautman spacetimes \cite{RobinsonTrautman:1960, RobinsonTrautman:1962, Stephanietal:book} represent deformations of spherically symmetric geometries and by radiating gravitational waves they asymptotically settle down to the Schwarzschild black hole. We hope to see a similar behavior when they are coupled to NE, this time settling down to the spherically symmetric NE configuration after possibly emitting gravitational and electromagnetic radiation. Such an approach was already successful in showing nonlinear stability of the Schwarzschild thin shell wormhole \cite{wormehole}. In \cite{Tahamtan-2016-NE}, we found very restricted Robinson–Trautman solutions (involving an electric charge only but no radiation) with NE sources for several specific models of the NE Lagrangian (both with and without a Maxwell limit). The solutions were generated from spherically symmetric ones. In all NE models we considered, the singularity of the electromagnetic field is resolved as in the static spherically symmetric cases. However, the models resolving the curvature singularity in spherically symmetric spacetimes could not be generalized to the Robinson–Trautman geometry using the approach of our previous paper \cite{Tahamtan-2016-NE}. Here we want to consider this problem in full generality and prove the impossibility of the generalization more rigorously.

{The Robinson--Trautman family of solutions has been used in situations more directly related to astrophysics as well. In \cite{Rezolla}, Rezzolla et al. showed an analytic explanation of an “antikick” appearing in numerical models of asymmetric binary black hole mergers by modelling the post-merger deformed horizon using Robinson--Trautman geometry and computed associated asymptotic momentum caused by directional gravitational radiation emission. Exact formulas for directional pattern of gravitational radiation in RT class show its direct relation to locations of higher horizon deformation. Evidently, RT family can provide useful idealized model for gravitational radiation from (slowly rotating) isolated sources, e.g., final stages of compact objects mergers. Subsequently, solutions coupled to electromagnetic fields can provide models containing electromagnetic counterpart to gravitational radiation. This is indeed the case for Robinson--Trautman--Maxwell solutions which in general contain both gravitational and electromagnetic radiation. Our aim is to analyze the situation when the RT geometry is coupled to NE. If general solutions do not contain electromagnetic radiation in this case this would provide potential clear observational difference compared to pure Maxwell case and suggest that if NE should play a role in the strong fields around compact objects there would be no strong electromagnetic counterpart to gravitational radiation in late merger stage. However, RT class is still not completely general to describe these situations in full and further study would be necessary to make predictions for general merger scenarios.} 

Previously, we found a solution containing exact gravitational waves for a specific NE Lagrangian (square root model) in Kundt class of geometries which is closely related to the RT class,  see \cite{Kundt-2016}. This provided hope that such general solutions can be found in RT class as well.

{{Although vacuum RT spacetimes asymptotically decay to Schwartzschild solution as shown analytically in \cite{Chru1,Chru2} and also RT--scalar field spacetimes settle down to spherically symmetric solutions \cite{Tahamtan-PRD-scalar}, the RT--Maxwell system was shown to be ill-posed in \cite{Lun94,Kozameh08}. This raises a question whether RT--NE system suffers from the same problem and thus might not be physically relevant and useful for stability analysis and other investigations. After briefly reviewing the study of ill-posedness of RT--Maxwell system we perform the analysis for the explicit RT--NE model obtained in this paper and show that it is well-posed. We plan to study well-posedness for general NE model in RT class in the future with the help of results established in \cite{wellpose}. }}

\section{Vacuum Robinson--Trautman metric}\label{RT-II}
\label{RTmetricsec}
The vacuum Robinson--Trautman spacetime can be described by the line element \cite{RobinsonTrautman:1960, RobinsonTrautman:1962, Stephanietal:book}
\begin{equation}\label{RTmetric}
\d s^2 = -2H\,\d u^2-\,2\,\d u\,\d r + \frac{r^2}{{P}^2}\,(\d x^{2} + \d y^{2}),
\end{equation}
where
\begin{equation}\label{RT-vacuum}
2H = K-2r(\,\ln{P})_{,u} -{2m/r}
\end{equation}
with $K=\Delta(\,\ln {P})$ and 
\begin{equation}\label{Laplace}
\Delta\equiv {P}^2(\partial_{xx}+\partial_{yy}).
\end{equation}
The metric generally contains two functions, ${\,{P}(u,x,y)\,}$ and ${\,m(u)\,}$. The function $m(u)$ might be set to a constant by a suitable coordinate transformation \cite{Stephanietal:book} and we assume this is the case for the coordinates of (\ref{RTmetric}). Einstein equations then reduce to a single nonlinear PDE --- the Robinson--Trautman equation
\begin{equation}
\Delta\Delta(\,\ln{P})+12\,m(\,\ln{P})_{,u}=0\,.
\label{RTeq}
\end{equation}
These spacetimes are then of algebraic type II.

As required by the definition of the Robinson--Trautman family the spacetime admits a geodesic, shearfree, twistfree and expanding null congruence generated by ${\boldl=\partial_r}$ with $r$ an affine parameter along this congruence, $u$~ the retarded time and $u=const$ null hypersurfaces. It shows the propagation of the tetrad components of the Weyl tensor along $u$, from null surface to null surface. Spatial coordinates $x,y$ span the transversal 2-space which has the Gaussian curvature (for ${r=1}$) 
\begin{equation}
{K}(u,x,y)\equiv\Delta(\,\ln{P})\,.
\label{RTGausscurvature}
\end{equation}
For general $r=const$ and $u=const$, the Gaussian curvature is ${{K}/r^2}$ so that, as ${r\to\infty}$, these 2-spaces become locally flat. As usual, we will assume that the transversal 2-spaces are compact and connected which leads to a subclass that contains the Schwarzschild solution (assuming a vanishing cosmological constant for simplicity) corresponding to ${K}=1$ (consistent with spherical symmetry). This subclass thus represents its generalization to a nonsymmetric dynamical situation.

To analyse the Robinson--Trautman equation \eqref{RTeq}, it is useful to introduce the following parametrization
\begin{equation}\label{PfP0}
{P}=f(u,x,y)\, {P}_{0}\,,
\end{equation}
where $f$ is a function on a 2-sphere $S^{2}$, corresponding to ${P}_0=1+\frac{1}{4}(x^2+y^2)$ (such choice gives ${K}=1$). By rigorously analysing equation (\ref{RTeq}) and substituting (\ref{PfP0}), Chru\'sciel \cite{Chru1,Chru2} proved that, for an arbitrary, sufficiently smooth initial data ${f(x,y,u_\textrm{i})}$ on an initial hypersurface ${u=u_\textrm{i}}$, Robinson--Trautman type~II vacuum spacetimes \eqref{RTmetric} exist globally for all ${u\ge u_\textrm{i}}$. Moreover, they asymptotically converge to the Schwarzschild--(anti-)de~Sitter metric with the corresponding mass $m$ and cosmological constant $\Lambda$ as ${u\to+\infty}$. This convergence is exponentially fast because $f$ behaves asymptotically as
\begin{equation}
f= \sum_{i,j\ge0} f_{i,j} u^j \e^{-2iu/m}\ ,
\end{equation}
where $f_{i,j}$ are smooth functions of the spatial coordinates $x,y$. For large retarded times~$u$, the function ${P}$ given by \eqref{PfP0} exponentially approaches ${P_0}$ which describes the corresponding spherically symmetric solution.

\section{Robinson--Trautman solution coupled to electromagnetism}

We consider the following action, describing an electromagnetic field in the form of nonlinear electrodynamics minimally coupled to gravity, 
\begin{equation}\label{action}
	S=\frac{1}{2}\int d^{4}x \sqrt{-g}\,\left[\mathbb{R}+\mathbb{L}(F,G)\right]\, ,
\end{equation}
where $\mathbb{R}$ is the Ricci scalar for the metric $g_{\mu \nu}$ (we use units in which $c=\hbar=8 \pi G=1$). $\mathbb{L}(F,G)$ is the Lagrangian of the nonlinear electromagnetic field which we assume to be an arbitrary function of the invariants $F$ and $G$ constructed from a closed Maxwell 2-form.
By varying the action (\ref{action}) with respect to the metric, we get Einstein equations
\begin{equation}\label{field equations}
	G^{\mu}{}_{\nu}=T^{\mu}{}_{\nu}\ .
\end{equation}

To keep the original form of the Robinson--Trautman spacetime, we assume the following metric 
\begin{equation}\label{RT-metric}
\d s^2 = -(2H+Q)\,\d u^2-\,2\,\d u\,\d r + \frac{R^2}{{P}^2}\,(\d x^{2} + \d y^{2}),
\end{equation} 
where $Q=Q(u,r,x,y)$ and $R=R(u,r)$ and subsequently we assume $u,r,x,y$ coordinate ordering. The metric function $H$ is presented in \eqref{RT-vacuum} where---to make our equations simpler--- we assume $m$ to be zero. However, one can always recover it as a $m/r$ term in $Q$.

In the next section we will find all the field equations for a general nonlinear electrodynamic Lagrangian. To do that, we will first obtain the modified Maxwell equations in the following subsection and then we will summarize Einstein equations.

\subsection{Modified Maxwell equations}
The Lagrangian of the nonlinear electrodynamics is generally supposed to be a scalar function of the invariants $F=F_{ab}F^{ab}$ and $G=F_{ab}F^{*ab}$. The electromagnetic fields obey the (generally) modified Maxwell (NE) field equations. The source--free nonlinear Maxwell equations are obtained in the standard way from the variational principle and they read
\begin{eqnarray}
F_{[\mu \nu,\lambda]}=0, \label{NE-MAX-TEN-2}\\
\left(\sqrt{-g}\,\mathbb{L}_{F}\,F^{\mu\nu}+\sqrt{-g}\,\mathbb{L}_{G}\,{}^{*}F^{\mu \nu}\right)_{,\mu}=0, \label{NE-MAX-TEN-1}
\end{eqnarray}
where we use the abbreviations $\mathbb{L}_F=\frac{\partial\,\mathbb{L}}{\partial F}$,  $\mathbb{L}_G=\frac{\partial\,\mathbb{L}}{\partial G}$, $\mathbb{L}_{FF}=\frac{\partial^{2} \mathbb{L}}{\partial^2 F}$, etc. We consider $F^{\mu \nu}$ to be the only fundamental variable as shown by Pleba\'{n}ski \cite{Plebanski}.  

The nonzero electromagnetic field components in the Robinson--Trautman class are $F_{ur}, F_{ux}, F_{uy}$ and $F_{xy}$. Since the metric \eqref{RT-metric} can accommodate only the outgoing rays aligned with the principle null direction, we assume that $F_{xr}, F_{yr}$ or $F^{ux}, F^{uy}$ are zero since they would otherwise correspond to rays in the opposite null direction (ingoing). This is related to fixing the initial conditions for the evolution of the Robinson--Trautman geometry. The electromagnetic field invariant is then 
\[F=F_{\mu \nu}\,F^{\mu \nu}=2\,(f_{m}-f_{e})\ ,\]
where
\[f_{m}=g^{xx}\,g^{yy}\,(F_{xy})^2,\,\,\,\,f_{e}=(F_{ur})^2\]
and the pseudoscalar invariant would be
\[G=F_{\mu \nu}\,{}^{*}F^{\mu \nu}=\frac{1}{2}\epsilon^{\mu \nu \alpha \beta}F_{\mu \nu}F_{\alpha \beta}\ ,\] 
where $\epsilon^{\mu \nu \alpha \beta}$ is the totally antisymmetric Levi-Civita tensor  with $\epsilon^{urxy}=\frac{-1}{\sqrt{-g}}$. Therefore, $G$ can be written in the following form
 \[G=-\frac{4}{\sqrt{-g}}\,F_{ur}F_{xy}.\]
 The duality relations between the electromagnetic field components take this form 
\begin{align} \label{dual} 
{}^{*}F^{xy}&=-\frac{F_{ur}}{\sqrt{-g}},&{}^{*}F^{ur}=-\frac{F_{xy}}{\sqrt{-g}}, \nonumber\\
{}^{*}F^{xr}&=\frac{F_{uy}}{\sqrt{-g}},&{}^{*}F^{yr}=\frac{F_{ux}}{\sqrt{-g}}. 
\end{align}
Maxwell equations (\ref{NE-MAX-TEN-2}) have the following components in our case
\begin{eqnarray}
\partial_{u}F_{xy}+\partial_{y}F_{ux}+\partial_{x}F_{yu}=0, \label{Second-NE-1}\\
\partial_{r}F_{xy}=0, \label{Second-NE-2} \\
\partial_{x}F_{ru}+\partial_{r}F_{ux}=0, \label{Second-NE-3} \\
\partial_{y}F_{ru}+\partial_{r}F_{uy}=0. \label{Second-NE-4}
\end{eqnarray}
It is possible to simplify the form of the electromagnetic field using the above equations. From (\ref{Second-NE-2}), we can integrate (we selected a convenient form to express the result)
\begin{equation}\label{Magnetic Field}
  F_{xy}=\frac{B(u,x,y)}{P(u,x,y)^2}
\end{equation}
and we also introduce a notation $F_{ur}=-E(u,r,x,y)$. Taking derivative of (\ref{Second-NE-1}) with respect to $r$, we obtain another useful relation
\begin{equation}\label{ux and uy}
(F_{ux})_{,yr}=(F_{uy})_{,xr}\ .
\end{equation}

The second modified Maxwell equation (\ref{NE-MAX-TEN-1}) for the metric (\ref{RT-metric}) has the following components (introducing the notation $\Xi=\sqrt{-g}\,\mathbb{L}_{F}$ and $\tilde{\Xi}=\sqrt{-g}\,\mathbb{L}_{G}$) 
\begin{align}
\left(\Xi\,F^{ru}+\tilde{\Xi}\,{}^{*}F^{ru}\right)_{,r}&=0\ ,\label{First-NEG-2}\\
\left(\Xi\,F^{rx}+\tilde{\Xi}\,{}^{*}F^{rx}\right)_{,r}+\left(\Xi\,F^{yx}+\tilde{\Xi}\,{}^{*}F^{yx}\right)_{,y}&=0\ , \label{First-NEG-3}\\
\left(\Xi\,F^{ry}+\tilde{\Xi}\,{}^{*}F^{ry}\right)_{,r}+\left(\Xi\,F^{xy}+\tilde{\Xi}\,{}^{*}F^{xy}\right)_{,x}&=0 \label{First-NEG-4}
\end{align}
and the last equation is
\begingroup\makeatletter\def\f@size{6}\check@mathfonts
\def\maketag@@@#1{\hbox{\m@th\normalfont#1}}
\begin{align}\label{First-NEG-1}
\left(\Xi\, F^{ur}+\tilde{\Xi}\,{}^{*}F^{ur}\right)_{,u}+\left(\Xi\, F^{xr}+\tilde{\Xi}\,{}^{*}F^{xr}\right)_{,x} 
+\left(\Xi\, F^{yr}+\tilde{\Xi}\,{}^{*}F^{yr}\right)_{,y}&=0. \nonumber \\
\end{align}
\endgroup
The above equations are not yet simplified--- to do that, we start with equation \eqref{First-NEG-2}. Using the duality relations between electromagnetic fields in \eqref{dual}, equation \eqref{First-NEG-2} can be written as 
\begin{equation}\label{First-NEG-22}
\Xi\,F_{ur}+\mathbb{L}_{G}F_{xy}=\tilde{C}(u,x,y).
\end{equation}
Applying now the above expression, we simplify equation (\ref{First-NEG-1}) further to obtain
 \begin{align}\label{First-NEG-11}
 \left(\mathbb{L}_F\,F_{ux}+\mathbb{L}_{G}F_{uy}\right)_{,x} 
 +\left(\mathbb{L}_F\,F_{uy}+\mathbb{L}_{G}F_{ux}\right)_{,y}-\left(\tilde{C}\right)_{,u}&=0 \nonumber\\ 
 \end{align}
 with $\tilde{C}$ independent of $r$. Taking derivative of (\ref{First-NEG-11}) with respect to $r$, we get the following constraint relation which will be useful later on to find the expressions for radiative fields,
 \[ \left(\mathbb{L}_F\,F_{ux}+\mathbb{L}_{G}F_{uy}\right)_{,xr}=-\left(\mathbb{L}_F\,F_{uy}+\mathbb{L}_{G}F_{ux}\right)_{,yr} .\]
 
The remaining Maxwell equations from the second set, namely (\ref{First-NEG-3}) and (\ref{First-NEG-4}), can be cast in the following form using our notation for electric and magnetic fields and the duality relation (\ref{dual})
 \begin{align}
 \left(\mathbb{L}_F\,F_{ux}+\mathbb{L}_{G}F_{uy}\right)_{,r}+\Omega_{,y}&=0, \label{First-NEG-33}\\
 \left(\mathbb{L}_F\,F_{uy}+\mathbb{L}_{G}F_{ux}\right)_{,r}-\Omega_{,x}&=0, \label{First-NEG-44}
 \end{align}
 where $\Omega=\left(\mathbb{L}_{G}\,E+\frac{\mathbb{L}_F\,B}{r^2}\right)$. Using (\ref{First-NEG-33}) and (\ref{First-NEG-44}), we can find the radiative fields $F_{ux}(u,r,x,y)$ and $F_{uy}(u,r,x,y)$ for any nonlinear electrodynamic Lagrangian 
 \begin{eqnarray}
 F_{ux}&=-\frac{\mathbb{L}_{F}(\int{\Omega_{,y}\,\d r}-\epsilon_{0})+\mathbb{L}_{G}(\int{\Omega_{,x}\,\d r}+\epsilon_{1})}{\mathbb{L}^2_{F}-\mathbb{L}^2_{G}}, \label{Fux-FG}\\
 F_{uy}&=\frac{\mathbb{L}_{F}(\int{\Omega_{,x}\,\d r}+\epsilon_{1})+\mathbb{L}_{G}(\int{\Omega_{,y}\,\d r}-\epsilon_{0})}{\mathbb{L}^2_{F}-\mathbb{L}^2_{G}}, \label{Fuy-FG}
 \end{eqnarray}
 where $\epsilon_{0}(u,x,y)$ and $\epsilon_{1}(u,x,y)$ are integration constants. Note that if we assume the electromagnetic fields to be independent from the transversal directions ($E(u,r)$ and $B(u)$) then (\ref{Fux-FG}) and (\ref{Fuy-FG}) imply 
 
 \begin{eqnarray}
 F_{ux}&=&\frac{\mathbb{L}_{F}\,\epsilon_{0}-\mathbb{L}_{G}\,\epsilon_{1}}{\mathbb{L}^2_{F}-\mathbb{L}^2_{G}}, \\
 F_{uy}&=&\frac{\mathbb{L}_{F}\,\epsilon_{1}-\mathbb{L}_{G}\,\epsilon_{0}}{\mathbb{L}^2_{F}-\mathbb{L}^2_{G}}. 
 \end{eqnarray}
Applying \eqref{Second-NE-3}  and \eqref{Second-NE-4} to this special case, we get radiative fields that are $r$-independent. However, analysing the above expressions, we conclude that this is only possible for a Lagrangian of the Maxwell type.


\subsection{Field equations}

After analysing the electromagnetic field equations we turn our attention to Einstein equations, both their geometrical and source parts.  

The electromagnetic energy momentum tensor is defined as 
\begin{equation}\label{energy-momentum-Maxwell}
T^{\mu}{}_{\nu}=\frac{1}{2}\{\delta ^{\mu}{}_{\nu}\,(\mathbb{L}-G\,\mathbb{L}_{G})-4\,(F_{\nu \lambda}F^{\mu \lambda})\mathbb{L}_F\}.
\end{equation}
Before proceeding to evaluate individual components of the energy momentum tensor we will use one component of Einstein equations, ${G^{u}}_{r}={T^{u}}_{r}$, to simplify the general metric (\ref{RT-metric}). Since $F_{xr}, F_{yr}$ are zero due to our general setting this Einstein equation simplifies considerably to
\begin{equation}
{G^{u}}_{r}=2\frac{R_{,rr}}{R}=0
\end{equation}
and gives $R$ that is linear in $r$ 
\[R=U_{1}(u)r+U_{2}(u)\ .\]
One can show that the above expression can be simplified into $R=r$ with a coordinate transformation, which we will assume from now on. It has been shown in \cite{Jacobson2007} that the above condition on energy momentum tensor components leads to this result for any static spherically symmetric spacetime of four or more dimensions whereas now we provided its generalization to the dynamical spacetime under consideration.
 
The nonzero components of the energy momentum tensor for electromagnetic fields are
\begin{eqnarray}
T^{u}{}_{u}&=&T^{r}{}_{r}=\frac{(\mathbb{L}-G\,\mathbb{L}_{G})}{2}+2\,f_{e}\mathbb{L}_F\ , \label{Tuu}\\
{T^{x}}_{x}&=&{T^{y}}_{y}=\frac{(\mathbb{L}-G\,\mathbb{L}_{G})}{2}-2\,f_{m}\mathbb{L}_F \label{Txx}
\end{eqnarray}
and, additionally, because we include radiation there are nonzero off--diagonal terms; i.e., 
\begin{eqnarray}
T^{r}{}_{u}&=&-2\,(F_{u \lambda}F^{r \lambda})\mathbb{L}_F \label{Tru} \\
&=&2\,\mathbb{L}_F\,\frac{P^2}{r^2}\left\{(F_{ux})^2+(F_{uy})^2\right\},  \nonumber \\
T^{r}{}_{x}&=&-2\,(F_{x \lambda}F^{r \lambda})\mathbb{L}_F \label{Trx}\\
&=&2\,\mathbb{L}_F\,\left\{F_{ur}F_{ux}+F_{uy}F_{xy}\frac{P^2}{r^2}\right\}, \nonumber\\
T^{r}{}_{y}&=&-2\,(F_{y \lambda}F^{r \lambda})\mathbb{L}_F \label{Try}\\
&=&2\,\mathbb{L}_F\,\left\{F_{ur}F_{uy}+F_{ux}F_{yx}\frac{P^2}{r^2}\right\} \nonumber
\end{eqnarray}
and also the following relations between the remaining energy momentum components,
\begin{eqnarray} \label{Txu-Tyu}
T^{x}{}_{u}=-\frac{P^2}{r^2}\,T^{r}{}_{x}\ ,\qquad
T^{y}{}_{u}=-\frac{P^2}{r^2}\, T^{r}{}_{y}\ .
\end{eqnarray}
 The geometrical part of the remaining Einstein field equations is given by the following expressions 
\begin{eqnarray}
G^{r}{}_{r}&=&G^{u}{}_{u}=\frac{Q_{,r}}{r}+\frac{Q}{r^2}, \label{Guu}\\
{G^{x}}_{x}&=&{G^{y}}_{y}=\frac{Q_{,rr}}{2}+\frac{Q_{,r}}{r} \label{Gxx}
\end{eqnarray}
and the off--diagonal terms are 
\begin{eqnarray}
G^{r}{}_{x}&=&-\frac{1}{2}\,Q_{,xr}, \label{Grx}\\
G^{r}{}_{y}&=&-\frac{1}{2}\,Q_{,yr} \label{Gry}
\end{eqnarray}
together with the relations similar to (\ref{Txu-Tyu}) 
 \begin{equation}
 G^{x}{}_{u}=-\frac{P^2}{r^2}\,G^{r}{}_{x},\qquad G^{y}{}_{u}=-\frac{P^2}{r^2}\,G^{r}{}_{y}.
 \end{equation}
 The last off--diagonal term---called the Robinson--Trautman equation in the vacuum case---is 
 \begin{align}
 {G^{r}}_{u}=\frac{-1}{2r^2}\Delta\left(K+Q\right)-\frac{1}{r}\left[(\ln P)_{,u}\left(rQ_{,r}-2Q\right)+Q_{,u}\right]\,. \label{RT-equation} \nonumber\\
 \end{align}
 All these field equations are obtained for the most general set up which can be used for any particular nonlinear electrodynamic Lagrangian. 

 \section{Consistency of the field equations}\label{consistency}
In this section, we check the consistency of Einstein equations with general NE in the Robinson--Trautman class described in section \ref{RT-II}. Using one of the Einstein equations \eqref{field equations},  $G^{u}{}_{u}=T^{u}{}_{u}$ with $G^{u}{}_{u}$ given by \eqref{Guu}, we obtain 
\begin{equation} \label{Equu}
(rQ)_{,r}=r^2\,T^{u}{}_{u}
\end{equation}
 Employing the expression for $T^{u}{}_{u}$ from \eqref{Tuu} and substituting the resulting equation in $G^{x}{}_{x}=T^{x}{}_{x}$ with the respective sides expressed using \eqref{Gxx} and \eqref{Txx}, we arrive at 
\begin{eqnarray}
G^2\,\mathbb{L}_{GG} \left(\frac{rE_{,r}-2E}{4E}\right)-G\,\mathbb{L}_{FG}\left(2rE_{,r}E+F\right)\nonumber\\
=-4E^2\mathbb{L}_{FF}\left(rE_{,r}E+\frac{2B^2}{r^4}\right)+\mathbb{L}_{F}\left(rE_{,r}E+2E^2\right),\nonumber\\
\end{eqnarray}
which can be alternatively expressed as
\begin{eqnarray} \label{Constraint}
rE_{,r}E \left[2F\,\mathbb{L}_{GG}+4E^2\left(\mathbb{L}_{FF}+\mathbb{L}_{GG}\right)-\left(2G\,\mathbb{L}_{FG}+\mathbb{L}_{F}\right)\right] \nonumber\\
+\frac{G^2}{2}\left[\mathbb{L}_{FF}-\mathbb{L}_{GG}\right]-F\,G\,\mathbb{L}_{FG}-2E^2\,\mathbb{L}_{F}=0.\nonumber\\
\end{eqnarray}
The above expression is equivalent to a component of the Maxwell equations (\ref{First-NEG-2}). If the Lagrangian would be in the form $\mathbb{L}(F)$ this equivalence is still satisfied and the form of the relation \eqref{Constraint} simplifies considerably 
\begin{eqnarray}\label{Constarint-LF}
\mathbb{L}_{FF}=\frac{\mathbb{L}_F\left(1+\zeta\right)}{4\,\left(\frac{B^2}{r^4}+E^2\,\zeta\right)},\nonumber \\
\end{eqnarray}
where $\zeta=\frac{r\,E_{,r}}{2E}$. One can arrive at this result by simply putting all $G$ derivatives of Lagrangian to zero in \eqref{Constraint}.
 
If we apply the same procedure for equation $G^{r}{}_{x}=T^{r}{}_{x}$ and substitute for $Q$ using \eqref{Equu}, we obtain an expression that, in the case of $\mathbb{L}(F)$ model, corresponds to a component of the Maxwell equation \eqref{Second-NE-3}. But for the $\mathbb{L}(F,G)$ model these two are not equivalent and equation $G^{r}{}_{x}=T^{r}{}_{x}$ has to be satisfied independently. We show the relevant computations in Appendix \ref{appendixA}.

\subsection{Example I: Maxwell theory} \label{Maxwell-Sec}
When studying nonlinear electrodynamics it is always worth it to first review the results for the linear theory, namely the Maxwell Lagrangian, in order to provide a comparison. The Maxwell theory corresponds to the Lagrangian $\mathbb{L}=-F$ which is also frequently considered a desirable weak-field limit (when $F$ and $G$ are small) of a general nonlinear Lagrangian. Note the overview  \cite{Stephanietal:book} of a complete solution for the Maxwell theory in the Robinson--Trautman class. The analysis therein was performed in the NP formalism. Here, we briefly review the Maxwell solution in the RT class using the tensorial formalism developed in preceding sections.

The independence of the Lagrangian on $G$ significantly simplifies most of the Maxwell equations. The magnetic field is still the same , $B(u,x,y)=P^2\,F_{xy}$, see \eqref{Magnetic Field}. To find the electric field, we use equation \eqref{First-NEG-22} and obtain the following expression for the electric field   
\begin{equation}\label{Electric Field-Maxwell}
E(u,r,x,y)=\frac{A(u,x,y)}{r^2},
\end{equation}
where $A(u,x,y)=P^2\,\tilde{C}$. To find the radiative fields, we use equations \eqref{Fux-FG} and \eqref{Fuy-FG},
\begin{eqnarray}
F_{ux}=\frac{B_{,y}}{r}-\epsilon_{0}, \label{By}\\
F_{uy}=-\frac{B_{,x}}{r}-\epsilon_{1}. \label{Bx}
\end{eqnarray}
From equations \eqref{First-NEG-11} and \eqref{ux and uy} we obtain the following relations
\begin{eqnarray}
\tilde{C}_{,u}-(\epsilon_{0})_{,x}-(\epsilon_{1})_{,y}=0, \label{Maxwell-Frist-4}\\
B_{,xx}+B_{,yy}=0.\label{LaplaceB}
\end{eqnarray}
Since we are interested in black-hole solutions we assume the transversal spaces to be compact and thus $\Delta B=0$ (see \eqref{LaplaceB}) means that $B$ should be a constant in $x$ and $y$ which further leads to $F_{ux}$ and $F_{uy}$ being $r$-independent due to (\ref{By}) and (\ref{Bx}). Using this result together with (\ref{Second-NE-3}) and (\ref{Second-NE-4}), we conclude that $E_{,x}=0$ and $E_{,y}=0$. Therefore we have $B(u)$, $E(u,r)$ and the Maxwell Lagrangian is independent of $x$ and $y$.

From \eqref{Electric Field-Maxwell} and the electromagnetic energy momentum tensor components (\eqref{Tuu}, \eqref{Txx}), the diagonal energy momentum components reduce to 
\begin{eqnarray}
T^{u}{}_{u}&=&T^{r}{}_{r}=-\left(\frac{A^2+B^2}{r^4}\right), \\
{T^{x}}_{x}&=&={T^{y}}_{y}\left(\frac{A^2+B^2}{r^4}\right) 
\end{eqnarray}
and the off--diagonal energy momentum components to
\begin{eqnarray}
T^{r}{}_{u}&=&-2\,\frac{P^2}{r^2}\left\{(F_{ux})^2+(F_{uy})^2\right\}, \label{Tru-Max}  \\
T^{r}{}_{x} &=&\frac{2}{r^2}\,\left\{A\,F_{ux}-B\,F_{uy}\right\}, \label{Trx-Max}\\
T^{r}{}_{y}&=&\frac{2}{r^2}\,\left\{A\,F_{uy}+B\,F_{ux}\right\}. \label{Try-Max}
\end{eqnarray}
Using $G^{u}{}_{u}=T^{u}{}_{u}$ and $G^{x}{}_{x}=T^{x}{}_{x}$, the metric function $Q$ takes the form
\begin{equation}
Q(u,r,x,y)=\frac{A(u)^2+B(u)^2}{r^2}-\frac{2\,m(u,x,y)}{r},
\end{equation} 
where we denoted the constant of integration $-2\,m$ to yield the proper Schwarzschild limit. Substituting the metric solution in $G^{r}{}_{x}-T^{r}{}_{x}=0$ and using \eqref{Grx} and \eqref{Trx-Max}, we get the following result 
\begin{equation}
m_{,x}=2\left(-A\,F_{ux}+B\,F_{uy}\right)
\end{equation}
and similarly for  $G^{r}{}_{y}-T^{r}{}_{y}$ with \eqref{Gry} and \eqref{Try-Max}, we obtain
\begin{equation}
m_{,y}=-2\left(A\,F_{uy}+B\,F_{ux}\right).
\end{equation}
Combining these equations with relation \eqref{Second-NE-1} for the fields and equation \eqref{Maxwell-Frist-4}, we can derive an expression for $m$
\[\Delta m=2P^2\,(B\,(F_{xy})_{,u}+A\,\tilde{C}_{,u}).\]
It can be written more explicitly as 
\begin{equation}\label{Max-m}
\Delta m=2(B\,B_,u-2(\ln P)_{,u}\,(A^2+B^2)+A\,A_{,u}).
\end{equation}
And finally, the last equation  $G^{r}{}_{u}=T^{r}{}_{u}$ (see \eqref{RT-equation} and \eqref{Tru-Max}) can be split into two equations for terms of different orders in $r$  
\begin{align}
\Delta K+12m\,(\ln P)_{,u}-4m_{,u}=&4P^2\left\{(\epsilon_{0})^2+(\epsilon_{1})^2\right\},\label{RT-Maxwell-eq}\\
\Delta m+4(\ln P)_{,u}(A^2+B^2)=&(A^2+B^2)_{,u}, \label{Max-ru-m}
\end{align}
where equation \eqref{Max-ru-m} is identical to \eqref{Max-m}. All equations for the Maxwell theory in the RT class are satisfied provided we solve the evolution equation (\ref{RT-Maxwell-eq}). It is clear that if one considers a vanishing electric charge (which corresponds to $A=0$) the resulting solution can still be a nontrivial radiative one and the same holds when $B=0$.

\subsection{Example II: Born--Infeld theory}
The Born--Infeld Lagrangian is one of the most attractive NE models which in the weak limit goes to the Maxwell case and in the strong regime goes to the square root model. This model was used in many different areas of research, from the flat spacetime with the aim to remove a point-charge singularity, to the string theory where it appears as a low--energy model. Also, there are many studies of its properties in GR, investigating, e.g., its stability, quasi normal modes, etc. and most recently the interaction with a scalar field \cite{Tahamtan-2020}.  In \cite{bicak1975}, the authors studied the Born--Infeld model in the flat spacetime  using the NP formalism and included the other electromagnetic invariant $G$. We aim to find an exact solution for the Born--Infeld model in a dynamical spacetime (the Robinson--Trautman class). We pay special attention to the radiative fields for BI model in this geometry. Note that most of the papers on nonlinear electrodynamics concern static or stationary spacetimes and studies of the dynamical behavior of this theory in exact form is absent in the literature. In this subsection, we use the field equations from the preceding sections to see whether such dynamical solutions exist in the RT class or not.

The Born--Infeld Lagrangian has the following form 
\begin{equation}\label{Born-Infeld}
\mathbb{L}(F,G)=4\,\beta^2\left(1-\sqrt{1+\frac{F}{2\,\beta^2}-\frac{G^2}{16\,\beta^4}}\right),
\end{equation} 
where $\beta$ is a constant which has the physical interpretation of a critical field
strength.
To solve the modified Maxwell equations for this specific Lagrangian, we start with \eqref{First-NEG-22} and obtain an expression for the electric field $E(u,r,x,y)$ in terms of the magnetic field $B(u,x,y)$ 
\begin{equation}\label{Electric-BI}
E(u,r,x,y)=\pm \frac{C\,\beta}{\sqrt{\beta^2\,r^4+B^2+C^2}},
\end{equation}
where $C=P^2\,\tilde{C}$ and $B$ \eqref{Magnetic Field} are functions of $u,x,y$. Compared to previous results concerning only the $\mathbb{L}(F)$ Lagrangian, the above electric field reproduces them when $B=0$, for example in paper \cite{Tahamtan-2016-NE} which investigated this model in the RT class. Curiously, when both $E$ and $B$ are nonzero the $\mathbb{L}(F)$ Lagrangian model produces an electric field that differs from the result obtained for $\mathbb{L}(F,G)$ and, moreover, the field is no longer regular at the origin. This illustrates the importance of including the invariant $G$ when considering the magnetic field as well.  The next step is to find the radiative fields using \eqref{Fux-FG} and \eqref{Fuy-FG} for the BI Lagrangian
\begin{eqnarray}
F_{ux}&=\Sigma \left[\Gamma(\int{\Omega_{,y}\,\d r}-\epsilon_{0})
-C\,B\left(\int{\Omega_{,x}\,\d r}+\epsilon_{1}\right)\right], \nonumber\\ \label{Fux-BI}\\
F_{uy}&=-\Sigma\left[\Gamma(\int{\Omega_{,x}\,\d r}+\epsilon_{1}) -C B\left(\int{\Omega_{,y}\,\d r}-\epsilon_{0}\right)\right] \label{Fuy-BI}, \nonumber\\
\end{eqnarray}
where 
\begin{eqnarray}
\Omega&=&-\frac{\beta \, B}{\sqrt{\beta^2\,r^4+B^2+C^2}}, \nonumber\\
\Gamma&=&\beta r^2\sqrt{\beta^2r^4+B^2+C^2}, \nonumber \\
\Sigma&=&\frac{\beta^2\,r^4+B^2}{B^2\,(\beta^2\,r^4-C^2)+\beta^2\,r^4(\beta^2\,r^4+C^2)}.\nonumber 
\end{eqnarray}
After calculating the electromagnetic fields we obtain the metric \eqref{RT-metric}, which is primarily determined by the metric function $Q$. Applying \eqref{Equu}, it takes the form
\begin{align}
Q(u,r,x,y)&=-\frac{m}{r}+\frac{2\beta^2}{3}r^2-\frac{2\beta}{r}\int \sqrt{\beta^2\,r^4+B^2+C^2},\nonumber\\
\end{align}
where $B$ , $C$ and now also $m$ are functions of $u,x,y$. Not surprisingly, the above solution is of the same form (apart from dependencies on $x,y$) as that found in \cite{Tahamtan-2016-NE} with the only difference being the presence of $B$. Since the electric fields found here and in \cite{Tahamtan-2016-NE} are similar, one would expect the similarity in the metric solution as well. However, one should as well note that we have also nontrivial radiative fields present in our current situation which was not the case in  \cite{Tahamtan-2016-NE}. With all the electromagnetic fields and the metric solution known, one should check whether the rest of the field equations are satisfied. Arriving at the equation $G^{r}{}_{x}=T^{r}{}_{x}$, we realize that it cannot be satisfied. Therefore, it is not possible to have a nontrivial electric and magnetic fields together with radiative fields for the Born--Infeld model in the RT class.

Note that with the assumption of vanishing radiative fields, one finds an exact solution in this theory with both electric and magnetic point charges, similar to \cite{BI-1986}. Moreover, it is clear that the functions $C$ and $B$ appearing in \eqref{Electric-BI} can only depend on $u$ then.

\subsection{Other models}

Besides the Born--Infeld model, the above result is valid for several other models of NE --- it is not possible to find consistent solutions for NE in the RT class with nontrivial radiative terms. Although some Einstein equations are equivalent to Maxwell equations (as shown at the beginning of section \ref{consistency}), the additional field equations containing radiative terms cannot be satisfied in the RT class.

\section{Robinson--Trautman solutions for $\mathbb{L}(F)$}

Due to the absence of nontrivial exact solutions of Einstein gravity coupled with NE Lagrangian when we assumed electric and magnetic fields in the RT class, in this section, we study the case when there exists only a magnetic (or an electric) field --- by this we mean non-null fields. Note that to consider ``pure magnetic field'' (the same would apply to ``pure electric field'') means the other invariant $G\sim \mathbf{E\cdot B}$ vanishes identically, so the form of Lagrangian is effectively $\mathbb{L}(F)$. We already found solutions for several models of NE for the $\mathbb{L}(F)$ Lagrangian in \cite{Tahamtan-2016-NE} in this spacetime while involving only an electric point charge without having electromagnetic radiation. In this section we concentrate on solutions for a magnetic charge. First, we find general formula for unknown metric function $Q$ in our metric ansatz \eqref{RT-metric} in the presence of ``magnetic field''. Then we evaluate conditions for finding regular black holes in the RT class and extend the solution to contain electromagnetic radiation as well. Furthermore, in the subsequent section, we use the same method to find an exact solution for ``electric field''.

\subsection{Magnetic field}

Recently, studying magnetic fields in NE became subject of substantial interest as a means to find regular black holes in static spherically symmetric spacetimes. By regular black holes we mean geometries lacking singularity at the center of a black hole determined by the presence of a horizon. Thus all scalar invariants are regular everywhere in the spacetime. The first regular black hole was introduced by Bardeen \cite{Bardeen-1968} as a solution generated by certain stress energy tensor without clear physical interpretation. More recently, number of models for regular black holes have been proposed together with physically motivated matter content needed for their explanation \cite{regulars-no-NE}.

The idea of constructing regular black holes by nonlinear electrodynamics as a source was introduced by Ay\'{o}n-Beato and Garc\'{\i}a \cite{Ayon-Beato1}. The same authors showed that the corresponding source of the Bardeen black hole can be associated with a specific model of nonlinear electrodynamics Lagrangian coupled to gravity \cite{Ayon-Beato-Bardeen}. Soon after that Bronnikov \cite{Bronnikov} proved a theorem which says that there is no spherically symmetric solution with a globally regular metric coupled to nonlinear electrodynamics satisfying correct weak--field limit and having nonzero electric charge. Note that all these solutions and most of the nonlinear electrodynamics models representing regular black hole spacetimes presented in literature are static spherically symmetric \cite{regular-with-NE}. Some of these regular black-hole solutions have been extended to stationary spacetimes \cite{stationary}.

Here, we are mainly interested in the possibility of having regular black holes in the RT class by utilizing suitable NE model as a source. As we already mentioned earlier based on \cite{Bronnikov}, regular black holes in static spherically symmetric (SSS) situations need only magnetic field. Here, we study the same situation although our spacetime is not static (not even stationary) nor spherically symmetric. But this approach may provide the possibility of extending those SSS solutions to the dynamical spacetime.

In general, no matter what model of NE is applied, the equation \eqref{Second-NE-2} always holds which means that the magnetic field $F_{xy}$ is independent of $r$. Since we are studying the case when there is only a magnetic field we put the electric field $E(u,r,x,y)$ and the radiative terms, $F_{ux}, F_{uy}$, to zero. The modified Maxwell equations are getting significantly shorter and simpler. 

Using \eqref{Second-NE-1} we conclude that $F_{xy}$ is $u$-independent and from \eqref{First-NEG-3} and \eqref{First-NEG-4} we obtain the following expression

\begin{equation}\label{Magnetic}
F_{xy}(x,y)=\frac{r^2\,C_{0}(u,r)}{P^2\mathbb{L}_F}.
\end{equation}
With no radiative terms, $F_{ux}=0$ and $ F_{uy}=0$, all the off--diagonal energy momentum tensor components are vanishing. Therefore, Einstein equations $G^{r}{}_{x}=T^{r}{}_{x}$ and  $G^{r}{}_{y}=T^{r}{}_{y}$ lead to $Q$ independent of $x,y$. Turning to the remaining Einstein equations (e.g., $G^{u}{}_{u}=T^{u}{}_{u}$) we have
\begin{equation} \label{magnetic-solution}
(rQ)_{,r}=r^2\mathbb{L}/2,
\end{equation}
which can be written in the following form 
\begin{equation} 
	Q(u,r)=\frac{1}{2r}\, \int r^2\mathbb{L}(F) \d r-\frac{m(u)}{r},  \nonumber
\end{equation}
where $"-m"$ is an integration constant. It is clear from the above equation that the Lagrangian should be $\mathbb{L}(u,r)$. Therefore $F$, the electromagnetic scalar invariant, should be $F(u,r)$. Since from the definition we have $F=\frac {2\,F^2_{xy}\,P^4}{r^4}$ then the function $P$ has necessarily a separated form, namely
\begin{equation} \label{separating-P}
P=\frac{q_m(u)}{\sqrt{F_{xy}}}\ .
\end{equation}
Then the electromagnetic scalar invariant is $F=\frac{2\,q_m^4}{r^4}$ and is obviously singular at $r=0$. The above expression for $P$ is also consistent with \eqref{Magnetic}.

Let us recall that the aim of using only the magnetic field here is to find regular black hole solutions defined by regularity of scalar curvature invariants such as Ricci and Kretschmann scalars. These two scalar quantities for our metric \eqref{RT-metric} can be expressed as 

\begin{eqnarray}
Kretschmann=
(Q_{,rr})^2+(\frac{2\,Q_{,r}}{r})^2+(\frac{2\,Q}{r^2})^2, \label{Kretschmann}\\
Ricci=-Q_{,rr}-\frac{4\,Q_{,r}}{r}-\frac{2\,Q}{r^2}. \label{Rici}
\end{eqnarray}
Let us note that the form of these two quantities is not changing even for more general case when $Q$ is a function of $x,y$ as well. For having regular solution the mass term $m/r$ must be excluded from $Q(u,r)$. The main equation in the Robinson--Trautman class which determines its dynamics is the so-called RT equation \eqref{RT-equation} (with ${T^{r}}_{u}=0$ now)
 \begin{align}
{G^{r}}_{u}=-\frac{1}{2r^2}\Delta K-\frac{1}{r}\left[(\ln P)_{,u}\left(rQ_{,r}-2Q\right)+Q_{,u}\right],  \nonumber\\
\end{align}
where $K(u,x,y)$ is $K=\Delta(\,\ln {P})$ as in the vacuum RT. Using \eqref{separating-P} it can be written as following 
\begin{equation}
\frac{q_m^4\,\kappa_{0}}{2r}+(\,\ln {q_m})_{,u} \{r{Q_{,r}}-2Q\}+{Q_{,u}}=0,
\end{equation}
where $\kappa_{0}=\Delta K\rvert_{q_{m}=1}$ is necessarily a constant. The solution for the above equation is 
\begin{equation}\label{Q-form}
Q=\frac{q_m^2}{2}\left(2\,f(r/q_m)-\frac{q_m\,\kappa_{0}}{r}\int q_m \d u\right)
\end{equation}
where $f$ is an arbitrary function. If we look at this solution for $Q$ with nonzero $\kappa_{0}$ (when $\kappa_0=0$ the solution reduces from algebraic type II to type D --- in effect leading to spherically symmetric case only) and plug it into scalar invariants \eqref{Rici} and \eqref{Kretschmann}, we see that it is not possible to have a regular solution. Eventhough there can be certain solutions for arbitrary nonlinear electrodynamics Lagrangian none of them can correspond to a regular black hole.

We can impose some assumptions on the magnetic field or the corresponding metric solution to see whether in some special cases it is possible to have regular solution.

One such assumption is that the so-called ``magnetic charge'', $q_{m}$, is a constant. Then from $F=\frac{2\,q^4_m}{r^4}$ the Lagrangian is now only $r$-dependent. So the metric function $Q$ \eqref{magnetic-solution} can be written in the following form
\begin{equation}\label{last-Q}
Q(u,r)=\mathcal{R}(r)-\frac{m(u)}{r}.
\end{equation}
For checking the regularity we again compute two scalar invariant quantities, \eqref{Kretschmann} and \eqref{Rici}. It is clear that parameter $m$ in \eqref{last-Q} has to vanish as previously. Let us assume that there exists some form of $\mathcal{R}(r)$ that makes the two scalar invariant quantities regular. But now we need to make sure that with this assumption the spacetime is still of type II, namely not spherically symmetric. From the main equation determining the dynamics of RT spacetime \eqref{RT-equation} we have
\begin{equation}
{G^{r}}_{u}=-\frac{1}{2r^2}\Delta\Delta(\,\ln {P})=0,
\end{equation}
which obviously shows that the spacetime is no longer type II but rather  type D only and thus spherically symmetric.

If we consider an electric charge instead of a magnetic one we will arrive at a form of $Q$ corresponding to \eqref{Q-form} and obtain type of solutions already discussed in \cite{Tahamtan-2016-NE}. These cannot give rise to spherically symmetric regular black hole solutions due to already discussed results of \cite{Bronnikov}.

The above negative results for regular black holes stem from two crucial facts. First, the necessity of assuming vanishing mass parameter $m$ to preserve regularity and from the absence of radiative terms that lead to restricted form of $Q$. This means that there are no relevant sources for nontrivial Gaussian curvature ($K\neq const.$) in RT equation \eqref{RT-equation}


\subsection{Magnetic charge and radiation}

In this section, we continue the study with same assumptions as in preceding subsection with the difference that we allow for radiative terms, i.e., $E=0$ but $F_{xy}$, $F_{ux}$, $F_{uy}$ are assumed nonzero.

From \eqref{Second-NE-3} and \eqref{Second-NE-4}, the radiative field components $F_{ux}, F_{uy}$ are independent of $r$ like for $F_{xy}$. Using the definition \eqref{Magnetic Field} and equations \eqref{First-NEG-33} and \eqref{First-NEG-44}, we get the following relations for these two radiative fields
\begin{equation}
F_{ux}=-\frac{(\mathbb{L}_{F}\, B)_{,y}}{(\mathbb{L}_{F} \,B)_{,x}}F_{uy}.
\end{equation}
Recalling that the electromagnetic scalar invariant is $F=\frac{2\,B^2}{r^4}$, the above equation will simplify to
 \begin{equation}\label{ux-uy}
F_{ux}=-\frac{B_{,y}}{B_{,x}}F_{uy}
\end{equation}
and from \eqref{First-NEG-11} and also \eqref{ux-uy}, we get
 \begin{equation}
(F_{ux})_{,x}=-(F_{uy})_{,y}.
\end{equation}
The solution \eqref{magnetic-solution} for Einstein equations from the preceding subsection still holds here and we can write 
\begin{equation}
\mathbb{L}=\frac{2}{r^2}(r\,Q)_{,r}.
\end{equation}
By taking "$x$" derivative from both sides and considering $F=\frac{2\,B^2}{r^4}$, we get 
\begin{equation}\label{BQ}
2\,\mathbb{L}_{F}\,{B_{,x}}{B}=r^2{(r\,Q)_{,rx}}
\end{equation}
and similarly by using $y$ derivative. From $G^{r}{}_{x}=T^{r}{}_{x}$ (\eqref{Grx}, \eqref{Trx}) and $G^{r}{}_{y}=T^{r}{}_{y}$ (\eqref{Gry}, \eqref{Try}), we further obtain
\begin{eqnarray}
-4\,B\mathbb{L}_F\,F_{uy}=r^2\,Q_{,xr}, \label{rx-1}\\
-4\,B\mathbb{L}_F\,F_{ux}=r^2\,Q_{,yr}. \label{rx-2}
\end{eqnarray}
With the help of \eqref{BQ} and the $y$-version of it we arrive at 
\begin{eqnarray}
-2\,{(r\,Q)_{,rx}}\frac{F_{uy}}{B_{,x}}=Q_{,xr}, \\
-2\,{(r\,Q)_{,ry}}\frac{F_{ux}}{B_{,y}}=Q_{,yr}.
\end{eqnarray}
Combining these two pairs of equations together and applying \eqref{ux-uy}, we get
\begin{eqnarray}
2\,r\,\frac{F_{ux}}{B_{,y}}(Q_{,x}-Q_{,y})-(Q_{,x}+Q_{,y})=C_{0},\\
2\,r\,\frac{F_{ux}}{B_{,y}}(Q_{,x}+Q_{,y})-(Q_{,x}-Q_{,y})=C_{1},
\end{eqnarray} 
where $C_{0}(u,x,y)$ and $C_{1}(u,x,y)$ are integration constants. By solving for $Q_{,x}$ and $Q_{,y}$ from the above equations and assuming $Q_{,xy}=Q_{,yx}$ one can show that $Q$ should be independent of $x,y$. By further using Einstein field equations, for example $G^{r}{}_{x}-T^{r}{}_{x}=0$ and $G^{r}{}_{y}-T^{r}{}_{y}=0$, we immediately conclude that the radiative fields must vanish. Therefore we end up with the case where only a magnetic field exists, like in the previous section. 

Note that we are generalizing the solutions based on the vacuum RT solution (containing $m(u)$). If we consider $"-\frac{m(u,x,y)}{r}"$ term in the vacuum RT solution with the new assumption that our Lagrangian is of the form $\mathbb{L}(u,r)$ then from Eqs \eqref{rx-1} and \eqref{rx-2} (instead of $Q$ one should put the vacuum metric function $H$ \eqref{RT-vacuum} now), we will see that only for the Maxwell case $\mathbb{L}_F=-1$ it is possible to have radiation consistent with $m(u,x,y)\neq 0$.

\subsection{Electric charge and Radiation}\label{NE-model-sol}

To compare with the preceding sections, we study the case when there exists only an electric field and electromagnetic radiation terms. In \cite{Tahamtan-2016-NE}, we found several solutions for different models of nonlinear electrodynamics for an electric point charge without considering electromagnetic radiation. This part would be a generalization of \cite{Tahamtan-2016-NE}. As usual we start with modified Maxwell equations, from \eqref{First-NEG-22} we have
\begin{equation}\label{Electric Field}
\mathbb{L}_F\,F_{ur}=\frac{A}{r^2},
\end{equation}
where $A(u,x,y)=P^2\,\tilde{C}$ and $\tilde{C}(u,x,y)$ is an integration constant. From \eqref{First-NEG-33} and \eqref{First-NEG-44} (note that we assume $F_{xy}=0$), we obtain
\begin{eqnarray}
\mathbb{L}_F\,F_{ux}=\tilde{C}_{1},\label{Rad1}\\
\mathbb{L}_F\,F_{uy}=\tilde{C}_{2},\label{Rad2}
\end{eqnarray}
where $\tilde{C}_{1}$ and $\tilde{C}_{2}$ are integration constants. From \eqref{Second-NE-1} we get
\begin{equation}
(F_{ux})_{,y}=(F_{uy})_{,x}.
\end{equation}
By checking the Einstein equations component $G^{u}{}_{u}=T^{u}{}_{u}$, we arrive at 
\begin{equation}
{(r\,Q)_{,r}}={r^2}\,T^{u}{}_{u} \nonumber
\end{equation}
which means that 
\begin{equation}\label{QE}
Q(u,r,x,y)=\frac{1}{r}\int r^2\,T^{u}{}_{u}\, \d r-\frac{2\,m(u,x,y)}{r}\ .
\end{equation}
Since we are looking for the possibility to have radiative terms, we check equations $G^{r}{}_{x}=T^{r}{}_{x}$ (see \eqref{Grx} and \eqref{Trx}) and $G^{r}{}_{y}=T^{r}{}_{y}$ (see \eqref{Gry} and \eqref{Try}) using \eqref{Electric Field} to arrive at the following expressions
\begin{eqnarray}
-4\,A\,F_{ux}=r^2\,Q_{,xr}, \\
-4\,A\,F_{uy}=r^2\,Q_{,yr}.
\end{eqnarray}
By taking $x,r$ and $y,r$ derivatives of $Q$ \eqref{QE}, namely $Q_{,xr}$ and $Q_{,yr}$, and putting them back to the above equations we find
\begin{equation}\label{Radiation}
-4\,A\,F_{ux}=r^3\,(T^{u}{}_{u})_{,x}-\int r^2\,(T^{u}{}_{u})_{,x}\, \d r+2\,m_{,x} .
\end{equation}
We can express the relevant energy momentum tensor component from \eqref{Tuu} 
\begin{eqnarray}
(T^{u}{}_{u})_{,x}&=&\frac{\mathbb{L}_{F}}{2}\,F_{,x}+\frac{2}{r^2}\left[A_{,x}F_{ur}+{A}(F_{ur})_{,x}\right],\\
F_{,x}&=&-4\,(F_{ur})\,(F_{ur})_{,x},
\end{eqnarray}
where $f_{e}=(F_{ur})^2$. Finally, we arrive at
\begin{equation}
(T^{u}{}_{u})_{,x}=\frac{2\,A_{,x}}{r^2}F_{ur}.
\end{equation}
Now \eqref{Radiation} would be (using \eqref{Rad1} and also \eqref{Electric Field})
\begin{equation}
\left(\frac{2\,\tilde{C}_{1}}{A_{,x}}\,r^2+r\right)\,\,F_{ur}=\int F_{ur}\, \d r-\frac{m_{,x}}{A_{,x}}
\end{equation}
and by $r$ derivative of the above equation while remembering $F_{ur}=-E$ we obtain
\begin{equation}\label{electric-with-radiation}
E(u,r,x,y)=-\frac{q_{e}(u,x,y)}{\left(\frac{2\,\tilde{C}_{1}}{A_{,x}}\,r+1\right)^2}
\end{equation}
and similarly from the same procedure, starting with $y,r$ derivatives before \eqref{Radiation}, we get alternative form of $E$. Comparing these two, we get this constraint equation
\[\frac{\tilde{C}_1}{A_{,x}}=\frac{\tilde{C}_2}{A_{,y}}.\]
From the expression for $E$ and using \eqref{Second-NE-3}, we can express radiative term in the following way 
\begin{equation}\label{Rad11}
	F_{ux}=- \int {E_{,x} \d r} +\tilde{\epsilon}_{0},
\end{equation}
where $\tilde{\epsilon}_{0}(u,x,y)$ is an integration constant. On the other hand by substituting $\mathbb{L}_F$ from \eqref{Electric Field} into \eqref{Rad1}, one expects that both expressions \eqref{Rad1} and \eqref{Rad11} for $F_{ux}$ are the same leading to equation
\begin{equation}
-\frac{A}{r^2\,E}(- \int {E_{,x} \d r} +\tilde{\epsilon}_{0})=\tilde{C}_{1}	.
\end{equation}

From this equality and the expression for the electric field \eqref{electric-with-radiation}, we get the following relations
\[q_e(u,x,y)=q_e(u), \quad \tilde{\epsilon}_{0}=\frac{q_e\,A^2_{,x}}{4\,\tilde{C}_{1}A}, \quad \tilde{C}_{1}=\frac{a\,A_{,x}}{\sqrt{A}}, \]
where $a(u)$ is an integration constant. Applying this procedure to $F_{uy}$ we find the relations   
\[\tilde{\epsilon}_{1}=\frac{q_e\,A^2_{,y}}{4\,\tilde{C}_{2}A}, \quad \tilde{C}_{2}=\frac{a\,A_{,y}}{\sqrt{A}}, \]
where $\tilde{\epsilon}_{1}(u,x,y)$ is an integration constant coming from \eqref{Second-NE-4} in a similar way to  $\tilde{\epsilon}_{0}$ in \eqref{Rad11}.

Using the above results we can re-express the electric field from \eqref{electric-with-radiation} in a simpler way 
\begin{equation}
	E(u,r,x,y)=-\frac{q_{e}}{\left(\frac{2\,a}{\sqrt{A}}\,r+1\right)^2}.
\end{equation}
One expects that this electric field corresponds to some specific NE Lagrangian. To find the Lagrangian, we first find $r$ in terms of $F$ ($F=-2\,E^2$ ) and substitute it in \eqref{Electric Field} to obtain the following expression
\begin{equation}\label{LF-new}
\mathbb{L}_{F}=\frac{4a^2}{q_{e}}\frac{1}{\left(-1+\left(-\frac{F}{2q_{e}^2}\right)^{1/4}\right)^2}.
\end{equation}
Such form clearly means that both $q_{e}$ and $a$ have to be constants and represent parameters of the Lagrangian model rather than, e.g., interpreting $q_{e}$ as a charge which is anyway not its proper interpretation.

Then we can integrate to obtain corresponding Lagrangian 
\begin{equation}\label{New-Lagrangian}
	\mathbb{L}_{New}=32q_e\,a^2\,\left[\frac{u^3+3u^2-4u-2}{2(1-u)}-3\ln{(1-u)}-\mathbb{L}_{0}\right],
\end{equation}
where $u=\sqrt{\sqrt{\frac{-F}{2q_e^2}}}$ and $\mathbb{L}_{0}$ is an integration constant. To have the exact Maxwell limit one should set $\mathbb{L}_{0}=-1$ but we consider $\mathbb{L}_{0}=0$ in the following calculations. Due to the logarithmic term in the Lagrangian we are limited to field strength corresponding to $u<1$ which can be fine-tuned using $q_{e}$ to accommodate high values of $F$. The plots of this Lagrangian together with Maxwell and Born--Infeld models are shown in Figure \ref{fig1}. Energy conditions for this new Lagrangian are investigated in Appendix \ref{appendixB}.
\begin{figure}
	\includegraphics[scale=0.57]{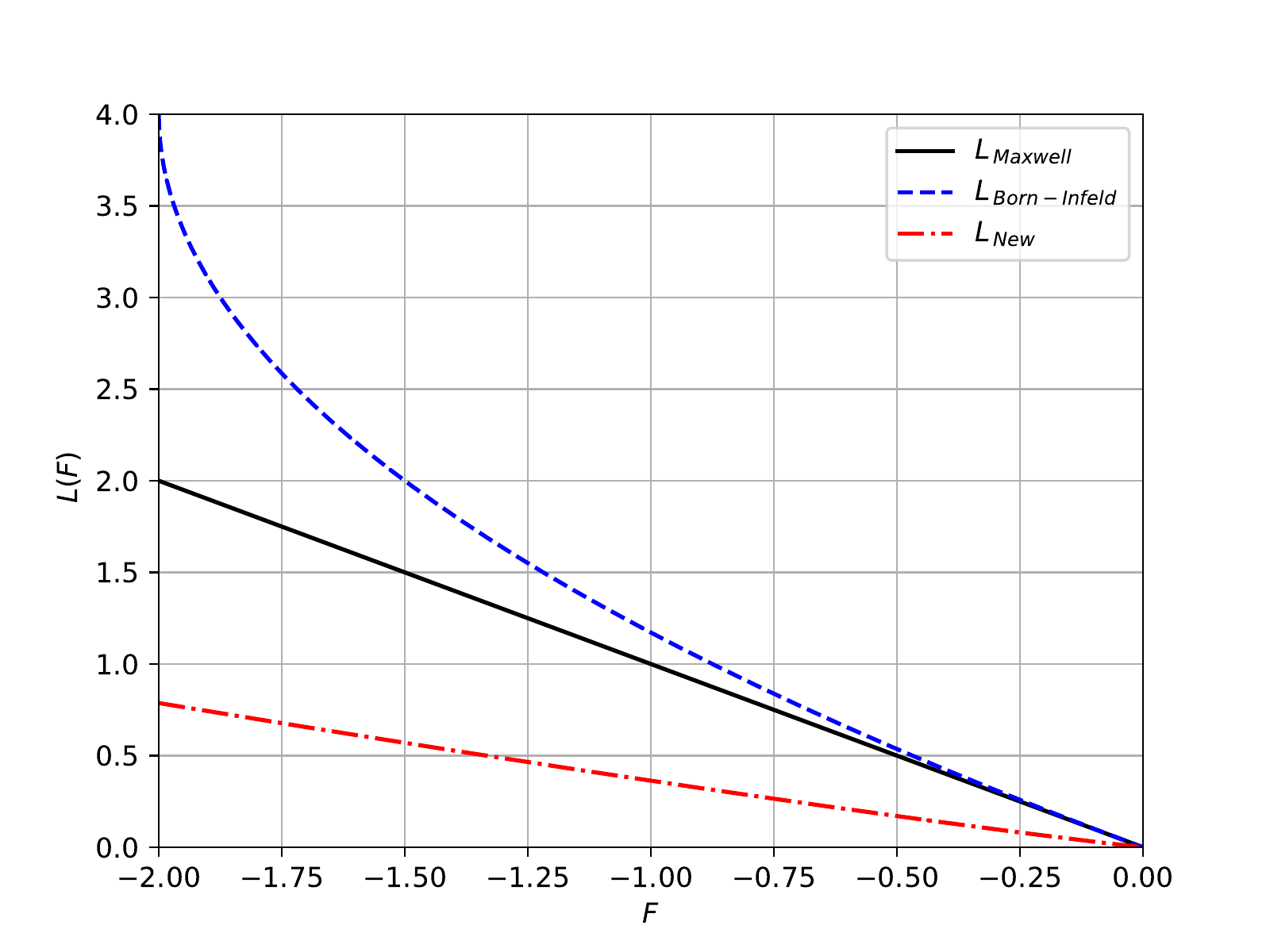}
	\caption{Plot of Maxwell, Born--Infeld and the new Lagrangian in terms of $F$. The constant parameters are set to $\beta=1$, $a=1$, $\mathbb{L}_{0}=-1$ and $q_e=-16$.}\label{fig1}
\end{figure}

Since the electric field and the Lagrangian are clear we can find the metric function by using $G^{u}{}_{u}=T^{u}{}_{u}$ and $G^{x}{}_{x}=T^{x}{}_{x}$ which leads to
\begin{eqnarray}
&&	Q(u,r,x,y)=\frac{\mu}{r}+\tilde{K}(u,x,y)-8q_e\,a\sqrt{A(u,x,y)}\,r\nonumber\\
&&+16\Lambda r^2+48\Lambda r^2\,\ln\left(\frac{2a\,r}{2a\,r+\sqrt{A(u,x,y)}}\right),\nonumber\\
\end{eqnarray}
where  $\tilde{K}=2q_eA$, $\Lambda=-\frac{a^2q_e}{3}$ and the modified mass term is
\[\mu(u,x,y)=-2m(u,x,y)+\frac{q_e}{3a}A^{\frac{3}{2}}\left(7-6\ln(2\sqrt{A})\right). \]

This solution has the following form asymptotically (for $r\to\infty$) 
\begin{equation}\label{Q-asymptotic}
  Q \rightarrow 16\Lambda \,r^2+\frac{3a \mu+2q_e\,A^{\frac{3}{2}}}{3ar}-\frac{q_eA^2}{4a^2\,r^2}+\frac{q_eA^{\frac{5}{2}}}{10\,a^3r^3}+O(\frac{1}{r^4})
\end{equation}

To satisfy further Einstein equations  $G^{r}{}_{y}=T^{r}{}_{y}$ and $G^{r}{}_{x}=T^{r}{}_{x}$ the function $\mu(u,x,y)$ should be $x,y$-independent ($\mu(u)$). The last equation to satisfy is  $G^{r}{}_{u}=T^{r}{}_{u}$ (expressed using \eqref{RT-equation} and \eqref{Tru}) which leads to these dynamical equations
\begin{eqnarray}
&&\Delta\left(K+\tilde{K}\right)-6\,\mu(\ln P)_{,u}+2\,\mu_{,u}=0, \label{RT-1}\\
&&\Delta A+ A^2\,\Delta(\ln A)+\frac{A^{\frac{3}{2}}}{aq_{e}}\,\left(2\tilde{K}\,(\ln P)_{,u}-\tilde{K}_{,u}\right)=0,\nonumber\label{RT-NE-2}\\
\end{eqnarray}

Provided we solve the above two coupled nonlinear parabolic type PDEs \eqref{RT-1} and \eqref{RT-NE-2} for unknown functions $P$ and $A$, we have a complete specific solution for our model. Such a solution is generally radiative (both in terms of electromagnetic and gravitational waves), has the Maxwell-type limit and contains a term mimicking the cosmological constant (see \eqref{Q-asymptotic}). The metric solution is singular at $r=0$ while the electric field is not, provided $a$ is positive. Strangely, this NE model seems to be a unique one compatible with general solutions containing electromagnetic radiation in the RT class.

If we assume from the beginning that the electric field has restricted dependence on coordinates ($E(u,r)$) then using \eqref{Second-NE-3} and \eqref{Second-NE-4}, we derive that $F_{ux}$ (and similarly $F_{uy}$) is $r$-independent and immediately from \eqref{Rad1} (or \eqref{Rad2}) one sees that $\mathbb{L}_F$ has to be a constant leading to Maxwell theory.

\section{Linear perturbations around stationary solution}

In the vacuum case the Robinson--Trautman spacetimes have well defined evolution to the future \cite{Chru1, Chru2}. However, this is not the case for Robinson--Trautman--Maxwell system. In \cite{Lun94}, Lun and Chow study linear perturbations around Reissner--Nordstr\"{o}m solution within the RT--Maxwell system and show their exponential divergence. This result was later confirmed by \cite{Kozameh08} leading to limited significance of RT--Maxwell solutions due to its ill-posedness.

First, we review the linear perturbation analysis of the Robinson--Trautman--Maxwell system using the equations derived in section \ref{Maxwell-Sec}. Subsequently, we apply the same procedure to RT--NE system based on the NE Lagrangian derived in section \ref{NE-model-sol}.

We will linearize the evolution equations using assumptions already made in \cite{Lun94}. The function $P$ determining the geometry of transversal two-spaces is considered in the following form, see \eqref{PfP0}, 
\begin{equation}\label{pertub-P}
	P=P_{0}(1+g)\ ,
\end{equation}
in which $P_0$ corresponds to spherical symmetry and the function $g(u,x,y)$ is a perturbation. Therefore it is possible to linearize the field equations by neglecting terms with higher orders of $g$ and its spatial derivatives.

In order to investigate the behavior of Robinson--Trautman--Maxwell system, first we derive two time evolution equations for perturbations around stationary solution using equations \eqref{RT-Maxwell-eq} and \eqref{Max-ru-m}. If we assume $A=B$ in \eqref{Max-ru-m} are constants equal to $q_0$ (following the assumption made in \cite{Lun94, Kozameh08}), then the equation would simplifies into
\begin{equation}\label{evol-1}
	\left(\ln P\right)_{,u}=-\frac{\Delta m}{8\,q_{0}^2}\ .  
\end{equation}
By substituting the above expression in \eqref{RT-Maxwell-eq} we arrive at the second time evolution equation, 
\begin{eqnarray}\label{evol-2}
	m_{,u}=\frac{\Delta K}{4}-\frac{3}{8\,q_{0}^2}\,\Delta m-P^2\left\{(\epsilon_{0})^2+(\epsilon_{1})^2\right\}\ .
\end{eqnarray}

Based on \eqref{pertub-P}, the following linearization holds
\[K=\Delta \ln P \simeq 1+2g+\tilde{\Delta}g\]
where $\tilde{\Delta}\equiv P^{2}_{0}(\partial_{xx}+\partial_{yy})$. For the subsequent analysis we assume that perturbations can be expanded into spherical harmonics and work with individual modes, thereby obtaining 
\[\tilde{\Delta}\,g=-l(l+1)g\ .\]
Similar to \ref{pertub-P}, we assume linear perturbation for $m$, i.e.,
\[m=m_{0}+\psi\]
and 
\[\tilde{\Delta}\,\psi=-l(l+1)\,\psi\ .\]
Neglecting all higher-order terms in $g$, $\psi$ and their derivatives, the equations \ref{evol-1} and \ref{evol-2} lead to this linearized system
\begin{eqnarray}\label{evol-1-2}
	\psi_{,u}&=&\frac{3\,l(l+1)\,m_{0}}{8\,q_{0}^2}\,\psi+\frac{l(l+1)\left[l(l+1)-2\right]}{4}\,g\ ,\nonumber \\
	g_{,u}&=&\frac{l(l+1)}{8\,q_{0}^2}\psi \ .
\end{eqnarray}
Such system provides exponential solution for both  $g$ and $\psi$ ($\sim \e^{\gamma u}$) determined by eigenvalues
\begin{equation}
	\gamma=\frac{l(l+1)}{16\,q_{0}^2}\left(3\,m \pm \sqrt{9m^2+8\,q_{0}^2\left[l(l+1)-2\right]} \right)\ .
\end{equation}
The solution contains two branches corresponding to eigenvalues of opposite signs for any $l \geq 1$, thus the linearized equations are unstable. For the negative sign, the limit $q_{0} \to 0$, goes to the vacuum linearized RT solution and this is the stable branch. The stability and well--posedness for RT class with Maxwell is discussed in detail in \cite{Lun94, Kozameh08}. These studies concluded that linearized RT with Maxwell is unstable and ill-posed. However, \cite{Lun94} observed using Lyapunov functional approach that spacetimes including past apparent horizon automatically correspond to the stable branch.

We will perform the same procedure for the only nonlinear electrodynamics radiative solution, \eqref{RT-1} and \eqref{RT-NE-2}, we have. First, we substitute $(\ln P)_{,u}$ from \eqref{RT-NE-2} to \eqref{RT-1} to have two time evolution equations as following 

\begin{eqnarray}
	&&\Delta\left(K+2q_e\,A\right)-6\,\mu(\ln P)_{,u}+2\,\mu_{,u}=0\ , \label{RT-11}\\
		&&\Delta\left(K+2q_e\,A\right)-3\,\mu(\ln A)_{,u}+2\,\mu_{,u}
		+\frac{3a\,\mu}{2\sqrt{A}}\times \nonumber\\&&
		\left[\frac{\Delta A}{A^2}+\Delta (\ln A)\right]=0\ . \label{RT-NE-22}
\end{eqnarray}

For studying linear perturbation , apart from \eqref{pertub-P}, we assume function $A$ to be 
\begin{equation}\label{pertub-A}
	A=A_{0}(1+\xi)\ .
\end{equation}
$A_0$ is a constant in spherically symmetric limit and 
\[\tilde{\Delta}\,\xi=-l(l+1)\,\xi\ .\]
Applying linear perturbation to equations \eqref{RT-11} and \eqref{RT-NE-22} while considering $\mu=-\mu_{0}$ as a constant, we obtain the following equations
\begin{eqnarray}\label{NE-evol-1-2}
	\xi_{,u}&=&2\,{\omega}\left\{\Pi\,\xi-\left[l(l+1)-2\right]\,g\right\}\ , \\\nonumber\\
	g_{,u}&=&{\omega}\left\{\sigma\,\xi-\left[l(l+1)-2\right]\,g\right\}\ ,
\end{eqnarray}
where the constants $\omega, \sigma, \Pi$ are ${\omega}=\frac{l(l+1)}{6\,\mu_{0}}$, $\sigma=2\,A_0\,q_e$ and  $\Pi=\sigma-\frac{3\,a\mu_{0}}{2}\left(\frac{1+A_{0}}{A^{\frac{3}{2}}_{0}}\right)$.

In this case the eigenvalues are
\begin{eqnarray}
	&&\gamma=\frac{l(l+1)}{12\,\mu_{0}^2}\left(2\,(1+\Pi)-l(l+1)\right.\nonumber\\
	&&\left.\pm \sqrt{\left(l(l+1)-2\right)^2+4\left(l(l+1)-2\right)(\Pi-2\,\sigma)+4\,\Pi^2} \right)\nonumber\\
\end{eqnarray}
and considering that  $q_e$ is negative while all the other constants are positive one can make both eigenvalues negative by adjusting the value of parameter $a$ controlling the strength of the Lagrangian. Thus the linearized equations are stable for any $l \geq 1$. Additionally, for fixed time and $l\to \infty$ the perturbations vanish as well. This specific RT--NE model is well-posed (since a norm on the initial data can be constructed that would bound norm of the solution at later times from above due to exponential decay ultimately leading to continuous dependence on initial data).

Surprisingly, there is a big difference in the stability of solutions between Robinson--Trautman--Maxwell and Robinson--Trautman with specific model of nonlinear electrodynamics. Although the nonlinear electrodynamics model found in the section \ref{NE-model-sol} has Maxwell limit but the corresponding solution in Robinson--Trautman class is well posed while Robinson--Trautman--Maxwell is not.

\section{Conclusion and discussion}

The main purpose of this paper was to derive conditions for the existence of NE solutions in the RT class that would contain electromagnetic radiation. Since general RT solutions represent nonlinear deformations of spherically symmetric situations, this would provide a potential basis for investigation of nonlinear stability of spherically symmetric NE solutions. This would be especially interesting for regular black--hole solutions sourced by NE. The presented results show that the Maxwell theory is the unique one providing general solutions including electromagnetic radiation while NE models suffer from serious restrictions--- with models featuring regular black holes affected in particular. The only partial exception to this is a NE model derived in section \ref{NE-model-sol} which provides a solution with an electric charge and radiation which has a Maxwell limit and involves a singular geometry but regular electromagnetic field.

Our results specifically mean that even the Born--Infeld model cannot support magnetic and electric charges accompanied by radiative terms in the general RT geometry. This model also offers an interesting justification for considering the $\mathbb{L}(F,G)$ Lagrangian instead of just $\mathbb{L}(F)$ when $B\neq 0$ and $E\neq 0$ are involved. Namely, the regularity of the electric field is destroyed for $\mathbb{L}(F)$.   

Further results for a magnetic charge only (facilitating an effective transition $\mathbb{L}(F,G) \to \mathbb{L}(F)$) with or without radiative terms explicitly show the impossibility of generalizing regular black--hole solutions to the RT class. This means one cannot hope to perform nonlinear stability analysis within this class but it also raises the question whether this means that such solutions are only limited to highly symmetric cases and thus do not represent astrophysically relevant situations arising from generic conditions. This represents additional complications for regular black--hole models which already suffer from a non-Maxwellian weak--field limit and, in the strong--field limit, both the value of their Lagrangian and energy momentum tensor attain constant values (this applies to, e.g., Lagrangians corresponding to the Bardeen and Hayward models).

{As noted in Introduction, RT geometry can be used for analysis of late stage of compact objects merger in certain scenarios. Viewed from this perspective, the clear lack of radiative solutions for RT--NE system in comparison with RT--Maxwell one suggests that one should not expect strong electromagnetic counterparts to gravitational radiation in these situations even when the sources of electromagnetic field are present if NE plays a role.}

{ Since it was previously shown that RT--Maxwell system is unstable when linearized around stationary solution we have applied such analysis to the specific NE model derived in section \ref{NE-model-sol} and concluded that this RT--NE system is perturbatively stable and well-posed. Thus it does not suffer from limited physical significance as in the case of RT--Maxwell.}

Naturally, one should pursue investigation in more general geometries to confirm the above results there. Especially spacetimes where radiation is not confined dominantly along one direction. This might prove to be essential for NE since general electromagnetic perturbations do not propagate along null directions here \cite{perlik2016}. Furthermore, the issue of well-posedness of general NE within the RT geometry can be tackled using results in  \cite{wellpose}. In the future we plan to clarify how the general RT solution for the specific model of NE settles to the stationary spacetime (a NE counterpart to Reissner--Nordstr\"{o}m solution).

\begin{acknowledgments}
The author is grateful to Otakar  Sv\'{\i}tek and David Hilditch for valuable comments and discussions and also thanks Martin \v{Z}ofka for carefully reading the paper.  It is our informal contribution to theoretical investigations related to gravitational waves within the LISA Consortium. This work was supported by the research grant GA\v{C}R 21-11268S.
\end{acknowledgments}

\appendix

\section{$\mathbb{L}(F)$ vs. $\mathbb{L}(F,G)$ model}\label{appendixA}

Here, we will consider the distinction between the $\mathbb{L}(F)$ and the $\mathbb{L}(F,G)$ models with respect to Einstein equations component $G^{r}{}_{x}=T^{r}{}_{x}$ and its relation to the Maxwell equations. This equation can be cast in the following form
\begin{equation} \label{Eqrx}
-\frac{1}{2}\,Q_{,xr}=2\,\mathbb{L}_F\,\left\{F_{ur}\,F_{ux}+F_{uy}\,\frac{B}{r^2}\right\}
\end{equation}
with $T^{r}{}_{x}$ and $G^{r}{}_{x}$ being given in \eqref{Trx} and \eqref{Grx}. From \eqref{Equu}, we can find 
\begin{equation}
Q_{,xr}=r\,(T^{u}{}_{u})_{,x}-\frac{1}{r^2}\int r^2\,(T^{u}{}_{u})_{,x} \d r
\end{equation}
 and by substituting the above expression into \eqref{Eqrx} we obtain
\begin{eqnarray}
r\,(T^{u}{}_{u})_{,x}-\frac{1}{r^2}\int r^2\,(T^{u}{}_{u})_{,x} \d r \nonumber \\
=-4\,\mathbb{L}_F\,\left\{F_{ur}\,F_{ux}+F_{uy}\,\frac{B}{r^2}\right\}.
\end{eqnarray}
To remove the integral we take $r$-derivative of this equation 

\begin{eqnarray}\label{Fux}
r^3\,(T^{u}{}_{u})_{,xr}+2r^2\,(T^{u}{}_{u})_{,x}=-4\left\{A\,F_{ux}+F_{uy}\,{\mathbb{L}_F\,B}\right\}_{,r}.\nonumber\\
\end{eqnarray}
Upon further simplification, using \eqref{Tuu}, we have the following 
\begin{eqnarray}
r^3\,(T^{u}{}_{u})_{,xr}+2r^2\,(T^{u}{}_{u})_{,x}=\nonumber\\ \mathbb{L}_F\left(r^2\,F_{,x}-\frac{8}{r^2}B\,B_{,x}\right)-\frac{4}{r^2}B^2\,F_{,x}\,\mathbb{L}_{FF}
\end{eqnarray}
 and by substituting the above expression into \eqref{Fux} we get
\begin{eqnarray}\label{Fux-2}
\mathbb{L}_F\left(r^2\,F_{,x}-\frac{4}{r^2}B\,B_{,x}\right)=-4\left(A\,F_{ux}\right)_{,r}
\end{eqnarray}
where we used \eqref{First-NEG-44} and \eqref{Electric Field}. With further straightforward simplification, we obtain a result equivalent to one component of the Maxwell equations \eqref{Second-NE-3}. Thus for $\mathbb{L}(F)$ it is straightforward to satisfy both the considered Einstein equations component and the Maxwell equations component since one is an integrability condition of the other.

If we repeat the same procedure for the $\mathbb{L}(F,G)$ model we obtain the following expression 
\begin{align}
  &F_{ux}=\\
  &\left(\frac{r^5}{8\,B}\right) \frac{r^2E\mathbb{L}_F\boxed{\left(E_{,x}+(F_{ux})_{,r}\right)}+B\mathbb{L}_G\left((F_{ux})_{,r}-E_{,x}\right)}{B\,r^2\mathbb{L}_{GG}\left(2E-rE_{,r}\right)+\mathbb{L}_{FG}\left(2B^2+r^5\,EE_{,r}\right)}. \nonumber
\end{align}
The presence of terms related to $G$-derivatives of Lagrangian makes the above equation completely independent from the Maxwell equations (unlike in the $\mathbb{L}(F)$ case) which results in a serious constraint on finding solutions to a coupled system. The final expression clearly significantly simplifies for the $\mathbb{L}(F)$ model and only the boxed terms survive leading to previous result.

\section{Energy conditions}\label{appendixB}
In this section, we review the energy conditions for our new Lagrangian \eqref{New-Lagrangian}, the other model of NE considered in this paper is Born--Infeld which was studied in great detail in literature. The essential condition for avoiding exotic matter is satisfying the weak energy condition (WEC) for any timelike vector $n^{\mu}$. The WEC is given by
\begin{equation}
	T^{\mu}{}_{\nu}\,n_{\mu}{} n^{\nu} \geq 0.
\end{equation}
For NE energy momentum tensor \eqref{energy-momentum-Maxwell} the above relation leads to the following expression
\begin{equation}
	-\frac{1}{2}(\mathbb{L}-G\,\mathbb{L}_{G})-2\,(F_{\nu \lambda}\,n^{\nu\,}F^{\mu \lambda} \,n_{\mu})\mathbb{L}_F \geq 0\,,
\end{equation}
which in the absence of $G$ (as is the case for solution considered in section \ref{NE-model-sol}) leads to the following observation. If both Lagrangian $\mathbb{L}$ and its derivative $\mathbb{L}_F$ are everywhere negative, WEC holds for arbitrary solution. This is not the case for our situation of Lagrangian \eqref{New-Lagrangian} since by considering constant $q_e$ to be negative, then $\mathbb{L}$ is positive (Figure \ref{fig1}) and $\mathbb{L}_F$ \eqref{LF-new} is negative. Therefore, one needs to analyze the situation for a given specific solution. Since the energy momentum tensor is not diagonal in our coordinate system the investigation is not straightforward. 

The time like vector in the most general form should be $(n^u,n^r,n^x,n^y)$ with $n_{\mu}n^{\mu}=-1$. For simplicity, let us consider the timelike vector with only $n^u$ component being nontrivial. For our non-diagonal metric anzats, \eqref{RT-metric}, the corresponding covector has form $(n_{u}, n_{r},0,0)$ and clearly $n_{r}=-n^u$. With these assumptions $n^u$ can be expressed as
\[n^u=\frac{1}{\sqrt{1-g_{uu}}}\]
where the denominator is positive and greater than $1$ in the exterior of the horizon. The WEC for such timelike vector, $n^{\mu}$, is
\begin{equation}
	T^{\mu}{}_{\nu}\,n_{\mu}{} n^{\nu}=T^{r}{}_{u}\,n_{r}{} n^{u}+T^{u}{}_{u}\,n_{u}{} n^{u} \geq 0.
\end{equation}
and using the explicit expressions for components of our timelike vector (covector) the WEC simplifies into

\begin{equation}\label{WEC}
\frac{T^{u}{}_{u}-T^{r}{}_{u}}{1-g_{uu}} -T^{u}{}_{u}\geq 0.
\end{equation}

\begin{figure}[t]
	\includegraphics[scale=0.57]{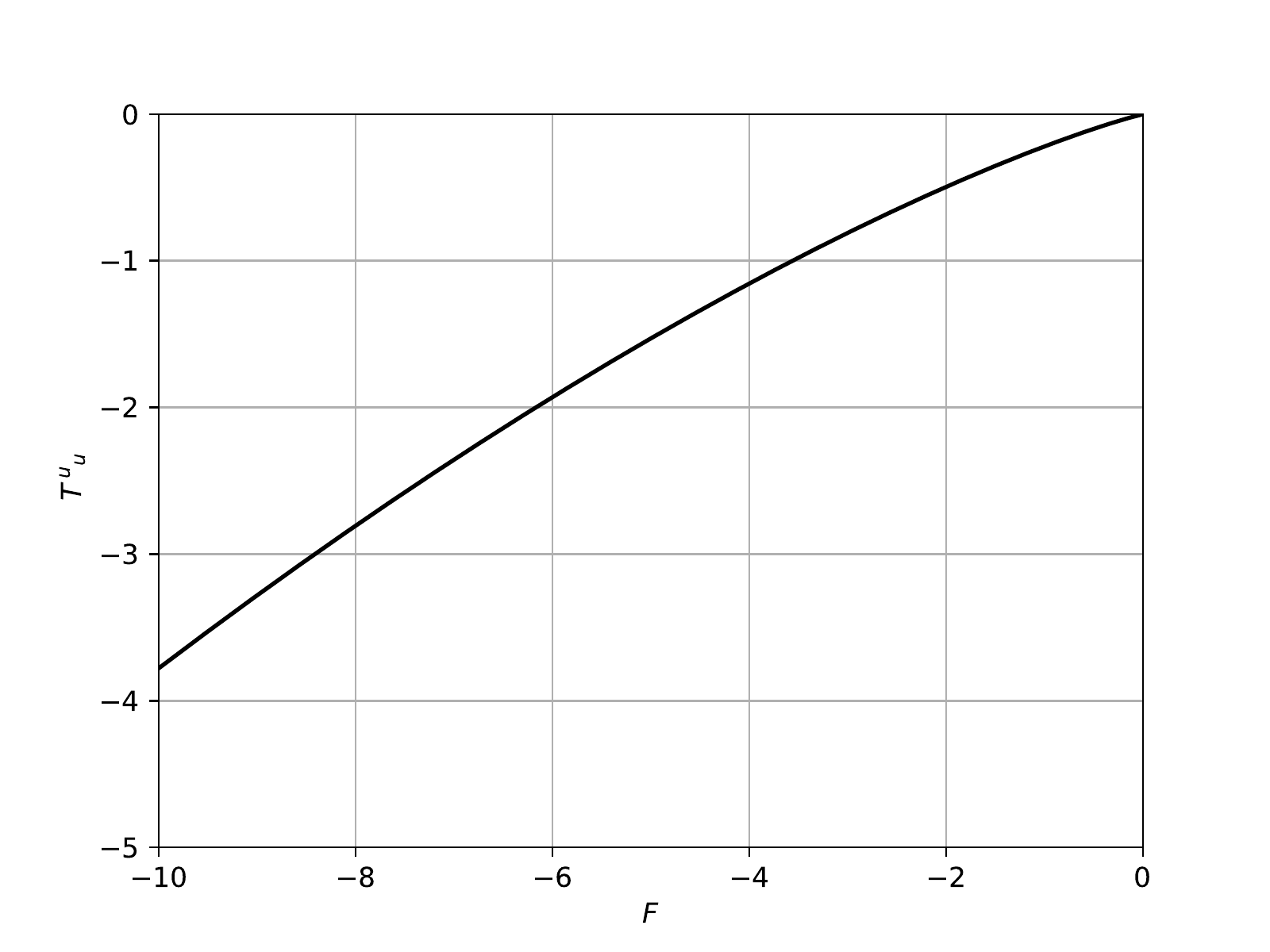}
	\caption{Plot for the energy momentum tensor component $T^{u}{}_{u}$ corresponding to the new Lagrangian \eqref{New-Lagrangian} in terms of $F$. The constant parameters are set to $a=1$, $\mathbb{L}_{0}=-1$ and $q_e=-16$.}\label{fig2}
\end{figure}

The energy momentum tensor component $T^{r}{}_{u}$, \eqref{Tru}, for the new Lagrangian \eqref{New-Lagrangian} is
\[T^{r}{}_{u}=\frac{2\,q_{e}P^2\,\,\left(\tilde{C}^2_{1}+\tilde{C}^2_{2}\right)}{A\,\left(\frac{2\,a}{\sqrt{A}}\,r+1\right)^2}\]
and considering negative $q_e$ the above expression is always negative. The component  $T^{u}{}_{u}$ is also negative as confirmed by its plot for specific values of constants in Figure:\ref{fig2} and this holds whenever $q_e$ is negative. Therefore, the WEC energy condition \eqref{WEC} is satisfied for this observer. Moreover, this result holds for arbitrary observer in the absence of radiative fields as well.

The Null Energy Condition (NEC) is satisfied for any null vector $k^{\mu}$ provided
\begin{equation}
	T^{\mu}{}_{\nu}\,k_{\mu}{} k^{\nu} \geq 0,
\end{equation}
which for our case means 
\[-2\,(F_{\nu \lambda}\,k^{\nu\,}F^{\mu \lambda} \,k_{\mu})\mathbb{L}_F \geq 0\]
and by assuming $q_e$ negative we see from \eqref{LF-new} that $\mathbb{L}_F$ is always negative. Therefore, NEC is automatically satisfied for the new Lagrangian.

\end{document}